\documentstyle[prd,aps,twocolumn]{revtex}
\begin{document}
\draft

\twocolumn[\hsize\textwidth\columnwidth\hsize\csname @twocolumnfalse\endcsname
\title{Toward Quantum Gravity I: Newton Gravitation Constant, Cosmological Constant, and Classical Tests}
\author{Heui-Seol Roh\thanks{e-mail: hroh@nature.skku.ac.kr}}
\address{BK21 Physics Research Division, Department of Physics, Sung Kyun Kwan University, Suwon 440-746, Republic of Korea}
\date{\today}
\maketitle

\begin{abstract}
This study toward quantum gravity (QG) introduces an $SU(N)$ gauge theory with the
$\Theta$ vacuum term as a trial theory. Newton gravitation constant $G_N$ is realized
as the effective coupling constant for a massive graviton, $G_N /\sqrt{2} = g_f
g_g^2/8 M_G^2 \approx 10^{-38} \ \textup{GeV}^{-2}$ with the gauge boson mass $M_G =
M_{Pl} \approx 10^{19}$ GeV, the gravitational coupling constant $g_g$, and the
gravitational factor $g_f$. This scheme postulates the effective cosmological constant
as the effective vacuum energy represented by massive gauge bosons, $\Lambda_e = 8 \pi
G_N M_G^4$, and provides a plausible explanation for the small cosmological constant
at the present epoch $\Lambda_0 \approx 10^{-84} \ \textup{GeV}^2$ and the large value
at the Planck epoch $\Lambda_{Pl} \approx 10^{38} \ \textup{GeV}^2$; the condensation
of the singlet gauge field $\langle \phi \rangle$ triggers the current anomaly and
subtracts the gauge boson mass, $M_G^2 = M_{Pl}^2 - g_f g_g^2 \langle \phi \rangle^2 =
g_f g_g^2 (A_{0}^2 - \langle \phi \rangle^2)$, as the vacuum energy. Relations among
QG, general relativity, and Newtonian mechanics are discussed.
\end{abstract}

]
\narrowtext

\section{Introduction}

Einstein's general theory for relativity \cite{Eins} is a presently accepted,
classical theory for gravitation and is tested in the weak field approximation.
However, general relativity or standard big bang theory based on general relativity
has outstanding problems: the singularity, cosmological constant or vacuum energy,
flat universe, baryon asymmetry, horizon problem, the large scale homogeneity and
isotropy of the universe, dark matter, galaxy formation, discrepancy between
astrophysical age and Hubble age, etc. These problems might be resolved and the
eventual unification of fundamental forces might be possible if quantum theory for
gravitation is found. General relativity is to be replaced by quantum theory for
gravitation in order to unify gravity with the other fundamental forces. Nevertheless,
there is so far no satisfactory quantum alternative for general relativity and
accordingly no quantum cosmology for standard big bang theory. The longstanding
problems in physics are thus abstracted as the quantization of gravity, the
cosmological constant problem, and the unification of gravitation with the other
fundamental forces. These problems are the motivations for quantum gravity (QG)
interested in this paper. According to recent experiments, BUMERANG-98 and MAXIMA-1
\cite{Jaff}, the universe is flat and there exists repulsive force represented by
non-zero vacuum energy, which plays dominant role in the universe expansion. In this
paper and subsequent paper \cite{Roh11}, the quantum and classical features of QG as a
gauge theory associated with a group $G$ are suggested to resolve the theoretical
problems in general relativity and are justified by the experimental facts even though
the group $G$ is exactly not known at present: the group chain is given by $G \supset
SU(2)_L \times U(1)_Y \times SU(3)_C$ where $G$, $SU(2)_L \times U(1)_Y$, and
$SU(3)_C$ groups are for gravitation, weak \cite{Glas}, and strong \cite{Frit}
interactions respectively. Specifically, an $SU(N)$ gauge theory as a trial theory
toward QG is introduced to resolve topics relevant for Newton gravitation constant
$G_N$ and the cosmological constant and to achieve the possible generation of all the
fundamental interactions in terms of dynamical spontaneous symmetry breaking (DSSB).
This paper also tries to illustrate classical tests verified by Einstein's general
relativity: quantum tests predicted by gauge theories, from the Planck scale
$10^{-33}$ cm to the universe scale $10^{28}$ cm, are discussed in subsequent paper
\cite{Roh11}. The present work is mainly restricted to the low, real dimensions of
spacetime without considering supersymmetry. This work is based on phenomenology below
the Planck scale: Newtonian mechanics, Einstein's general relativity, and quantum
gauge theories.

Einstein successfully developed the classical field equation known as general
relativity theory for gravity \cite{Eins}, which arises naturally from a systematic
and explicit account of the geometrical structure of space and time as dynamic
constituents of nature, affected by matter or energy. The standard cosmology based on
the Einstein's general theory of relativity is the hot big bang model. This model is
supplemented by the success of the grand unified theory (GUT) \cite{Geor} based on
quantum gauge theory, that is able to show how the universe evolved since the initial
big bang. A logical consequence of the GUT phase transition is inflation which
promises to resolve some of its outstanding problems, flatness and horizon problems.
Quantum gauge theory which is the marriage of quantum theory and gauge symmetry has
been indispensable in providing a precise description of the microcosm. Three
fundamental forces except gravity are described by quantum gauge theory; the $U(1)_e$
gauge theory for electromagnetic force is one of the most successful theory in physics
at the moment and the $SU(2)_L \times U(1)_Y$ gauge theory for weak force \cite{Glas}
and the $SU(3)_C$ gauge theory for strong force \cite{Frit} are non-Abelian extensions
of the $U(1)_e$ gauge theory. These two fundamental theories, general relativity and
quantum gauge theory, can together cover the energy regime from the elementary
particle to the universe in describing the behavior of particles. However, the
outstanding problem of unifying gravity and the other fundamental forces in nature is
that general relativity theory is difficult to formulate as a renormalizable gauge
theory. All the known covariant or canonical quantization methods are not yet
successful in quantizing the gravitational field, even though they work well for the
other physical fields. There is no distinct connection between Einstein's general
relativity on which the standard cosmology is based and gauge theory on which GUT is
based. The incompatibility of the two modern theories, general relativity and gauge
theory, is thus the biggest obstacle to the unification of the two theories into one
theoretical framework; as known widely, the unification of the two theories has been
one of the greatest challenge in physics. There are usually two directions toward
quantum gravitation theory or the unification of fundamental forces: superstring
theory and Kaluza-Klein theory in the higher dimensions and the Planck scale.
Superstring theory is considered to be one of the most promising candidate in the
unification of forces but there has been no known compactification method to break
down to the real, low energy world and no clear answer to how superstring theory
solves the cosmological constant problem.

Another longstanding problem in cosmology is so-called the cosmological constant
problem since it has been introduced by Einstein \cite{Eins1}. The cosmological
constant is regarded as one of the most fundamental physical entities and various
attempts, which include supersymmetry, anthropic consideration, adjustment mechanics,
changing gravity, and quantum gravity, have been made but none of them is completely
able to resolve the problem \cite{Wein0}. The effective cosmological constant is very
small or nearly zero according to the astronomical observation while the several
theories demand that the cosmological constant should be relatively large
\cite{Wein0}. The discrepancy in the cosmological constant or vacuum energy density
$10^{122} \ \textup{GeV}^4$ between the experimental observation and the theoretical
expectation are, on the one hand, more enormous than any other physical quantities. On
the other hand, according to recent experiments, BUMERANG-98 and MAXIMA-1 \cite{Jaff},
the universe is flat and there exists repulsive force represented by non-zero vacuum
energy, which plays dominant role in the universe expansion. This phenomena is
contradicted with general relativity which is accepted, classical theory for
gravitation. In this context, it is quite natural to develop quantum gauge theory to
overcome these theoretical problems and to explain experimental observations.

An $SU(N)$ gauge theory with the $\Theta$ vacum toward QG is in this paper introduced
as a trial theory to resolve topics relevant for
Newton gravitation constant $G_N$ and the cosmological constant
and illustrates the possible classical and quantum tests.
Newtonian gravitation constant $G_N$, which is regarded as one of fundamental constants, is
postulated as the effective coupling constant like Fermi constant $G_F$ in weak interaction \cite{Glas}.
The dimensionality of $G_N$, which is the obstacle of renormalization, is overcome by
introducing the gravitational coupling constant $\alpha_g$ and the massive gauge boson (graviton)
with the Planck mass $M_G = M_{Pl}$.
The cosmological constant is also defined in terms of the gauge boson mass by
the effective cosmological constant $\Lambda_e = 8 \pi G_N M_G^4$, which varies from
the huge value $\Lambda_e = \Lambda_{Pl} = 10^{76} \ \textup{GeV}^2$ at the Planck scale
to the measured, small value $\Lambda_e = \Lambda_0 = 10^{-84} \ \textup{GeV}^2$ at the present universe.
In order to test QG, classical tests predicted by
general relativity and verified by observations are here considered and
quantum tests predicted by a quantum gauge theory are discussed in the subsequent paper \cite{Roh11}.
Newtonian mechanics is postulated as quadrupole interactions of QG. Among classical tests, the deflection
of light by the Sun and the precession of the perihelia of the orbits of the inner
planets are the typical tests carried out both in empty space and in gravitational
fields that are approximately static and spherically symmetric \cite{Wein}. The energy
per mass, which exhibits the classical tests, obtained from Einstein's general
relativity in the weak field limit are therefore examined from the viewpoints of QG as
a gauge theory.

This paper is organized as follows. In Section II, an $SU(N)$ gauge theory toward QG
is suggested without considering supersymetry and then DSSB is introduced. Section III
describes the origin of Newton gravitation constant and the dual Meissner effect
presented as a plausible reason for no detection of gravitational waves in this
scheme. Section IV deals with the resolution of the cosmological constant problem as
the vacuum energy density.  In Section V, classical tests for gravity are discussed
from the potential QG point of view, in which Newtonian mechanics is postulated as a
quadrupole interaction of QG. Section VI addresses comparison between QG and general
relativity.  Section VII is devoted to conclusions.

\section{Toward Quantum Gravity}

An $SU(N)$ gauge invariant Lagrangian density with the $\Theta$ vacuum term is, without taking into account supersymmetry, introduced
toward QG as a trial theory even though the exact group $G$ for gravity is not unveiled.
DSSB triggered by the $\Theta$ term is adopted to generate gauge boson mass and fermion mass.
Natural units with $\hbar = c = k_B = 1$ are preferred for convenience throughout this paper unless otherwise specified.

The gauge invariant Lagrangian density is, in four vector notation, given by
\begin{equation}
\label{qchr}
{\cal L} = - \frac{1}{2} Tr  G_{\mu \nu} G^{\mu \nu}
+ \sum_{i=1}  \bar \psi_i i \gamma^\mu D_\mu \psi_i
\end{equation}
where the subscript $i$ stands for the classes of pointlike spinors, $\psi$ for the
spinor, and $D_\mu = \partial_\mu - i g_g A_\mu$ for the covariant derivative with the
gravitational coupling constant $g_g$. Particles carry the local charges and the gauge
fields are denoted by $A_{\mu} = \sum_{a=0} A^a_{\mu} \lambda^a /2$ with matrices
$\lambda^a$, $a = 0,.., (N^2-1)$. The field strength tensor is given by $G_{\mu \nu} =
\partial_\mu A_\nu - \partial_\nu A_\mu - i g_g [A_\mu, A_\nu]$. A current anomaly
\cite{Adle} is taken into account to show DSSB in analogy with the axial current
anomaly, which is linked to the $\Theta$ vacuum in QCD as a gauge theory
\cite{Hoof2}. The bare $\Theta$ term is added as a single, additional
nonperturbative term to the Lagrangian density (\ref{qchr})
\begin{equation}
\label{thet}
{\cal L}_{QG} = {\cal L}_{P} + \Theta \frac{g_g^2}{16 \pi^2} Tr G^{\mu \nu} \tilde G_{\mu \nu},
\end{equation}
where ${\cal L}_{P}$ is the perturbative Lagrangian density (\ref{qchr}), $G^{\mu \nu}$
is the field strength tensor, and $\tilde G_{\mu \nu}$ is the dual of the field strength tensor.
Since the $G \tilde G$ term is a total derivative, it does not affect the perturbative aspects of the theory.

DSSB consists of two simultaneous mechanisms; the first mechanism is the explicit
symmetry breaking of gauge symmetry, which is represented by the gravitational factor
$g_f$ and the gravitational coupling constant $g_g$, and the second mechanism is the
spontaneous symmetry breaking of gauge fields, which is represented by the
condensation of gravitational singlet gauge fields.  Gauge fields are generally
decomposed by charge nonsinglet-singlet on the one hand and by even-odd discrete
symmetries on the other hand: they have dual properties in charge and discrete
symmetries. Four singlet gauge boson interactions in (\ref{thet}), apart from
nonsinglet gauge bosons, are parameterized by the $SU(N)$ symmetric scalar potential:
\begin{equation}
\label{higs}
V_e (\phi) = V_0 + \mu^2 \phi^2 + \lambda \phi^4
\end{equation}
which is the typical potential with $\mu^2 < 0$ and $\lambda > 0$
for spontaneous symmetry breaking. The first term of the right
hand side corresponds to the vacuum energy density representing
the zero-point energy by even parity singlets. The odd-parity
vacuum field $\phi$ is shifted by an invariant quantity $\langle
\phi \rangle$, which satisfies
\begin{equation}
\label{higs1}
\langle \phi \rangle^2 = \phi_0^2 + \phi_1^2 +  \cdot \cdot \cdot + \phi_{N}^2
\end{equation}
with the condensation of odd-parity singlet gauge bosons: $\langle \phi \rangle = (\frac{- \mu^2}{2
\lambda})^{1/2}$. DSSB is relevant for the surface term $\Theta \frac{g_g^2}{16 \pi^2}
Tr G^{\mu \nu} \tilde G_{\mu \nu}$, which explicitly breaks down the $SU(N)$ gauge
symmetry for quantum gravity through the condensation of odd-parity singlet gauge
bosons. The $\Theta$ can be assigned by an dynamic parameter by
\begin{equation}
\label{thev}
\Theta = 10^{-61} \ \rho_G /\rho_m
\end{equation}
with the matter energy density $\rho_m$ and the vacuum energy density $\rho_G = M_G^4$.

\section{Newton Gravitation Constant}

The longstanding problems of both quantum gauge theory for gravity and the cosmological
constant problem might be resolved through DSSB.
Although the exact group is not known at a moment, the features of quantum gravity as a gauge theory may be suggested and
the origin of Newton gravitation constant $G_N$ may be discussed.
The group chain $G \supset SU(2)_L \times U(1)_Y \times SU(3)_C$ and the effective coupling constant chain
$G_N \supset G_F \times G_R$ are introduced \cite{Glas,Roh3,Frit}.
In the following, Newton gravitation constant and graviton mass, coupling constants
for fundamental forces, and the dual Meissner effect are addressed.

\subsection{Newton Gravitation Constant and Graviton Mass}

Gravitational interactions are generated by the emission and absorption of vector
bosons with spin $1$, gravitons, rather than tensor bosons with spin $2$. Gravitons
are the analogs of photons for electromagnetic force and gluons for color force.
Newton gravitation constant $G_N$ is cast in a form that can be directly compared to
Fermi coupling constant $G_F$ of electroweak interactions. $G_N$ replaces graviton
propagation $\sqrt{2} g_f g_g^2 / 8 (k^2 - M_G^2 )$ with the graviton mass $M_G$ and,
in contrast to the dimensionless coupling constant $g_g$ and the gravitational factor
$g_f$, has the dimension of inverse energy square. Contrary to the photon, the
graviton must be massive, otherwise it would have been directly produced in the
gravitational decays. It, in fact, turns out that the graviton has the Planck mass
$M_G \simeq M_{Pl} = 1.22 \times 10^{19}$ GeV.

The gravitational interaction amplitude below the Planck energy is of the form
\begin{equation}
\label{prgrr}
{\cal M} = - \frac{g_f g_g^2}{4} J^{\mu} \frac{1} {k^2 - M_G^2}  J_\mu^\dagger
= \sqrt{2} G_N J^{\mu} J_{\mu}^\dagger
\end{equation}
where ${\cal M}$ is the product of two universal current densities
and $g_f$ is the gravitational factor,
which is defined by $g_f = \frac{1}{4} (g_3^\dagger \lambda^a g_1) (g_2^\dagger \lambda_a g_4)$
with the gravitational charge fields, $g_i$ with $i= 1 \sim4$,
in analogy with the color factor $c_f$ in QCD.
The current density $J^\mu$ will be discussed with relation to the cosmological
constant and the baryon asymmetry in the following sections.
If $k^2 << M_G^2$, the effective gravitational coupling becomes
\begin{equation}
\label{glprr}
\frac{G_N}{\sqrt{2}} = - \frac{g_f g_g^2}{8 (k^2 - M_G^2 )} \simeq
\frac{g_f g_g^2}{8 M_G^2} \simeq 10^{-38} \ \textup{GeV}^{-2}
\end{equation}
and the gravitational currents interact essentially at a point.
That is, in the low momentum transfer, the propagation between the currents disappears.

Gravitational gauge field has the Planck mass at the phase transition:
\begin{equation}
\label{grmsr}
M_G \approx M_{Pl} \approx 10^{19} \ \textup{GeV}
\end{equation}
which is reduced to a smaller value due to the condensation of the
singlet graviton. Note that the conventional relation $G_N = 1
/M_{Pl}^2$ is adjusted to $G_N \simeq \sqrt{2} g_f g_g^2 /8
M_{Pl}^2$ in (\ref{glprr}). The gauge boson mass below the Planck
energy can be cast by
\begin{equation}
\label{gama} M_G^2 = M_{Pl}^2 - g_f g_g^2 \langle \phi \rangle^2 = g_f g_g^2
[A_{0}^2 - \langle \phi \rangle^2]
\end{equation}
with the even parity singlet gauge boson $A_{0}$, the odd-parity singlet gauge boson
condensation $\langle \phi \rangle$, the gravitational charge factor $g_f$, and the coupling
constant $g_g$. The gravitational factor $g_f$ used in (\ref{gama}) is the symmetric
factor for a gauge boson with even parity and the asymmetric factor for a gauge boson
with odd parity. This process makes the breaking of discrete symmetries P, C, T, and
CP. The effective vacuum energy density $V_e (\bar \phi)$ in (\ref{higs}) is connected
with the gauge boson mass $M_G$ by $V_e = M_G^4$: $V_0 = M_{Pl}^4, \ \mu^2 = -2 g_f g_g^2
M_{Pl}^2, \ \lambda = g_f^2 g_g^4$. The gauge boson condensation subtracting the gauge boson
mass is related to the subtraction of the zero-point energy in the system. Newton
gravitation constant at low momentum transfer thus can be related to the graviton mass
by expression (\ref{glprr}). The gauge boson mass must be identical to the inverse of
the screening length, that is, $M_G = 1/ l_{Pl} \simeq G_N^{1/2}$. A graviton thus has
the enormous mass $M_G \approx 10^{19}$ GeV and its propagation is prevented from
observation for the low energy graviton; this is known as the dual Meissner effect.
This is the reason why gravitational force is phenomenologically so weak compared to
the other forces and gravitational wave is not detected.

The above description is illustrated more rigorously by considering the Yukawa potential due to the massive graviton;
the Coulomb potential for the energetic gauge boson is screened by the mass of the gauge boson and consequently the effective gravitation coupling is very weak
at the macroscopic scale.
The Yukawa potential associated with the massive gauge boson is, in the static limit $E \rightarrow 0$,
\begin{equation}
\label{yumur}
V (r) = \sqrt{\frac{g_f g_g^2}{4 \pi}} \frac{e^{-M_G (r - l_{Pl})}}{r}
\end{equation}
which represents the short range interaction for the low energy graviton.

QG as a gauge theory with a certain group G has gauge bosons,
gravitons, whose interactions depend on their energies and masses.
The graviton with the energy higher than the Planck mass
propagates like massless particle, the graviton with the energy
equal to the Planck mass shows the Newtonian interaction, and the
graviton with the energy lower than the Planck mass propagates
only inside the Planck scale. $G_N$ thus makes two types of
potentials in gravitational coupling; the Yukawa potential for the
graviton with the energy lower than the Planck mass ($E < M_{Pl}$)
and the constant potential for the graviton with the energy equal
to the Planck mass ($E \rightarrow M_{Pl}$). This implies that
there are generally two kinds of potentials for massive gauge
bosons due to the boundary condition; one is the constant
gravitational potential for $E \rightarrow M_G$ and the other is
the Yukawa type gravitational potential for $E < M_G$. The
graviton with the energy lower than the Planck mass has the Yukawa
potential $\propto e^{-M_G^{e} (r - l_{Pl})}$ with the Planck
length $l_{Pl} = 1/M_{Pl}$ and the effective mass $M_G^{e} = i k_G
= (M_G^2 - E^2)^{1/2}$, the graviton with the energy higher than
the Planck mass has outgoing propagation $\propto e^{i k r}$ with
$k = (E^2 - M_G^2)^{1/2}$, the graviton with energy equal to the
gauge boson mass ($E \rightarrow M_G$) has the constant potential.
This suggests that the effective interaction of the graviton with
the energy below the Planck energy is proportional to $G_N J_\mu
J^{\mu \dagger}$ in the low momentum transfer. Below the Planck
energy, a phase transition takes place from a large group $G$ to
the $SU(2)_L \times U(1)_Y \times SU(3)_C$ group, where the $SU(2)_L \times U(1)_Y$ denotes
the weak isospin symmetry and the $SU(3)_C$ denotes the color
symmetry, via a unification group $H$, that
is, $G \supset H \supset SU(2)_L \times U(1)_Y \times SU(3)_C$.
Massive gravitons may be connected to
intermediate vector bosons and photons at the electroweak scale
and to gluons and photons at the strong scale.

\subsection{Coupling Constants for Fundamental Forces}

Out of four fundamental forces in nature, the most familiar forces
are gravity and electromagnetism which are experienced in the
macroscopic scale. Strong interactions are limited in range to
about $10^{-13}$ cm and are insignificant even at the scale of the
atom $10^{-8}$ cm, but play an important role in the binding the
nucleus. Weak interactions with an even shorter scale $(\leq
10^{-15} \ \textup{cm})$ are so weak that they do not bind
anything, but they do play an important role in weak decay
processes. Such local interactions of strong and weak forces are
connected with their massive gauge bosons; for example, weak force
is restricted by the intermediate vector boson mass $M_W$ with the
effective coupling constant $G_F = \sqrt{2} i_f g_i^2/8 M_G^2$.
Newton gravitation constant due to the massive graviton is the
origin of all the effective coupling constants \cite{Glas,Roh3};
the effective coupling constant chain is $G_N \supset G_F \times
G_R$ with Fermi weak constant $G_F \approx 10^{-5} \
\textup{GeV}^{-2}$ and the effective strong coupling constant $G_R
= \sqrt{2} c_f g_s^2/8 M_G^2 \approx 10 \ \textup{GeV}^{-2}$.
Strengths of four forces are roughly in orders of magnitude $10,
10^{-2}, 10^{-5}$, and $10^{-40}$ for strong, electromagnetic,
weak, and gravitational forces respectively. The difference in
strength is more than a factor of $G_R/G_N \approx 10^{39}$
between strong and gravitational interactions. From the ratio of
$G_F/G_N \approx 10^{33}$, electroweak force is $10^{33}$ times
stronger than gravitational force. In gravitational interactions,
it is predicted that the typical cross section at a temperature $T
= 1$ GeV is $\sigma \simeq G_N^2 T^2 \approx 10^{-70} \
\textup{m}^2$ and the typical lifetime for a particle with the
mass $1$ GeV is $\tau = 1/\Gamma \simeq 1/G_N^2 m^5 \approx
10^{50}$ years. Similarly, in the proton decay $p \rightarrow
\pi^0 + e^+$ at the energy $E << M_{Pl}$, the proton would have
much longer lifetime than $10^{30}$ years if the decay process is
gravitational. Features of fundamental interactions are summarized
in Table \ref{fefu}.

A phase transition at the Planck scale creates massive gravitons and massless gauge bosons like
the Higgs mechanism of electroweak interactions \cite{Higg};
there is mixing between gravitation currents so as to produce massive gravitons and massless gauge bosons
as Nambu-Goldstone (NG) bosons \cite{Namb}.
Massive gravitons produce the Yukawa gravitational potential with Newton
gravitation constant outside the Planck scale for the graviton with the energy $E < M_{Pl}$.
Massless gauge bosons (photons) produce the Coulomb potential in the static limit and have a weaker coupling
constant than the electromagnetic coupling constant $\alpha_e =1/137$ for the photon.
There is therefore the possibility that massless gauge bosons (photons) with the
energy $E_\gamma \approx 10^{19}$ GeV $\approx 10^{42}$ Hz responsible for the $U(1)$ gauge theory were
left as relics of phase transition at the Planck era.

One possibility as a candidate of quantum gravity is an $SU(3)_G$ gauge theory
under the assumption of no supersymmetry and higher dimensions.
The phase transition of $SU(3)_G \rightarrow SU(2) \times U(1) \rightarrow U(1)_c$ takes place at the Planck energy
just as $SU(3)_I \rightarrow SU(2)_L \times U(1)_Y \rightarrow U(1)_e$ for weak
force \cite{Roh2,Glas} and $SU(3)_C \rightarrow SU(2)_N \times U(1)_Z \rightarrow U(1)_f$ for strong force \cite{Roh3}.
In this case, the groups for all the fundamental forces hold the $SU(3)$ gauge symmetry as an essentially analogous dynamics although
each fundamental force dominates at different energy;
the gauge group hierarchy is $SU(3)_G \supset SU(3)_I \supset SU(3)_C$ and
the effective coupling constant hierarchy is $G_N \supset G_F \supset G_R$.
At higher energies, weakly interacting massive particles are relatively dominant but at lower energies, strongly interacting massive particles
are relatively dominant.
In terms of the renormalization group analysis, the considerable coupling constant for gravitational interactions
$\alpha_g \simeq 0.019$ is extracted from the weak and strong coupling constant $\alpha_i = \alpha_s \simeq 0.12$ around the unification energy $10^2$ GeV
when three charges and six flavors are assumed.
The massless gauge boson for the $U(1)_c$ gauge theory might have the coupling constant $\alpha_g/8 \simeq 1/410$ for attractive interactions,
which is about one third of the electromagnetic coupling constant $\alpha_e = 1/137$ for the $U(1)_e$ gauge theory.
Whatever the group $G$ is, a test for this possibility will thus be the search of the massless gauge boson (photon)
with an extremely high frequency $10^{42}$ Hz.

\subsection{Dual Meissner Effect in Gravitational Waves}

General relativity shows the existence of gravitational waves but
they have not been detected so far. The plausible reason why
gravitational waves have not been observed is more focused on in
this part. DSSB is undergoing through the condensation of the
singlet graviton in the evolution stage of the universe so as to
produce the gravitational superconducting state in the vacuum.
Below the Planck energy, the condensation of the singlet gauge
field increases as the universe expands and particles make the
Cooper pairs in the superconducting state. The essential point is
that the gauge field acquires a vacuum expectation value and the
resulting phase transition is still underway in the present
universe.

In this case, the difficulty in the detection of gravitational
waves is easily overcome by the dual Meissner effect in
gravitational waves, which is analogous to the Meissner effect in
the electric superconductivity \cite{Meis}: the exclusion of the
gravitational electric field in the superconducting state outside
the Planck scale. The evolution of the universe makes DSSB during
which the condensation of the singlet gauge field proceeds, the
Cooper pair of particles is formed, and flux tube is made by the
gravitational electric field. The gravitational electric field is
therefore excluded in the gravitational superconducting state due
to the dual Meissner effect; this is interpreted by the fact that
the graviton with the energy lower than the Planck mass becomes
massive. If the Cooper pairs due to the gravitational magnetic
field in the vacuum are made, the exclusion of the gravitational
electric field is obvious in the present universe. The extremely
massive graviton therefore provides the reason why gravitational
wave has not been detected and has the effective weak coupling for
gravity $G_N$. This argument is dually analogous to the screening
mechanism of magnetic fields in the presence of electric
superconductor \cite{Aitc}. More discussion on the dual Meissner
effect is addressed in the topics of fermion mass generation
and classical tests of QG. On the other hand, the massless gauge
boson created during DSSB at the Planck epoch has the enormous
frequency $10^{42}$ Hz as described above.

\section{Cosmological Constant}

The outstanding problem of the cosmological constant may be
resolved from the view point of gauge theory for gravity, in which
the cosmological constant is regarded as the vacuum energy
represented by the gauge boson mass.
The constraint of the flat universe, $\Omega - 1 = 10^{-61}$, is required by quantum gauge
theory and is confirmed by experiments BUMERANG-98 and MAXIMA-1 \cite{Jaff}.
The vacuum represented by the
massive gauge bosons is quantized by the maximum wavevector mode
$N_R = i/(\Omega -1)^{1/2} \approx 10^{30}$ and the total gauge
boson number $N_G = 4 \pi N_R^3/3 \approx 10^{91}$; the
gauge boson mass $M_G$ is quantized by the maximum wavevector mode $N_R$.
DSSB takes place through the condensation of singlet gauge bosons leading to the
current anomaly and decreases the effective cosmological constant.
The effective cosmological constant is here defined by $\Lambda_e
= 8 \pi G_N M_G^4$: $\Lambda_e \approx M_{Pl}^2 \approx 10^{38} \
\textup{GeV}^2$ at the Planck epoch and $\Lambda_e = \Lambda_0 = 3
H_0^2 \approx 10^{-84} \ \textup{GeV}^2$ at the present epoch. The
effective cosmological constant $\Lambda_e = \Lambda_0 = \Lambda
(t=t_0)$ defined at the present epoch $t=t_0$ is a number with
unit $\textup{cm}^{-2}$, which is independent of particle mass and
Newton gravitation constant $G_N$, and the cosmological
interaction occurs even in the absence of any matter at all.

In the following, the cosmological constant problem, the
resolution of the cosmological constant, and the modification of general relativity
are described.

\subsection{The Cosmological Constant Problem}

The universal cosmological constant $\Lambda$ can be inserted in Einstein's
field equation for gravity without destroying the general covariance:
\begin{equation}
\label{eisc}
R_{\mu \nu} - \frac{1}{2} g_{\mu \nu} R = - 8 \pi G_N \tilde T_{\mu \nu} = - 8 \pi G_N T_{\mu \nu} + g_{\mu \nu} \Lambda
\end{equation}
where $R_{\mu \nu}$ denotes the Ricci tensor, $R = R^\nu_\nu$ the scalar curvature,
$G_N$ Newton gravitation universal constant, $T_{\mu \nu}$ the energy
momentum tensor with its trace $T = T^\mu_\mu$, $\tilde T_{\mu \nu}$ the total energy-momentum tensor,
and $\Lambda$ the bare cosmological constant.
From Einstein's field equation the effective cosmological constant takes the form
\begin{equation}
\label{comp}
\Lambda_{e} = \Lambda + 8 \pi G_N \langle \rho_m \rangle
\end{equation}
where $- \langle \rho_m \rangle g_{\mu \nu} = \langle T_{\mu \nu} \rangle $ is the energy-momentum tensor in a vacuum according to Lorentz invariance.
The total effective vacuum energy is expressed by
\begin{math}
\langle \rho_m \rangle_{e} =  \langle \rho_m \rangle + \Lambda /(8 \pi G_N)  .
\end{math}
Measurements of cosmological redshifts as a function of distance provide the upper bound on
$\Lambda_{e} < H_0^2$ or $\langle \rho_m \rangle_{e} < 8.07 \times 10^{-47} \ \textup{GeV}^4$ with the Hubble constant $H_0$ \cite{Hubb}.
If the vacuum energy density $\langle \rho_m \rangle$ is approximated by
\begin{math}
\langle \rho_m \rangle \approx \Lambda_{cut}^4 /(16 \pi^2)
\end{math}
with the wave number cutoff $\Lambda_{cut}$,
$\langle \rho_m \rangle = 2 \times 10^{71} \ \textup{GeV}^4$ in the case of $ \Lambda_{cut} = (8 \pi G_N)^{-1/2}$.
Since $\langle \rho_m \rangle_{e}$ is less than $10^{-47} \ \textup{GeV}^4$ two terms, $\langle \rho_m \rangle$ and $\Lambda / (8 \pi G_N)$, must cancel to
better than $118$ decimal places.
This is so-called the cosmological constant problem \cite{Wein0}.

The cosmological constant is also related to the vacuum energy density $V (\bar \phi)$ \cite{Lind} described
by a constant scalar field $\bar \phi$.
When the scalar field appears, the change in the vacuum energy density
$V (\bar \phi)$ enters into Einstein's field equation
\begin{equation}
\label{eist}
R_{\mu \nu} - 1/2 g_{\mu \nu} R = - 8 \pi G_N \tilde T_{\mu \nu} = - 8 \pi G_N (T_{\mu \nu} - g_{\mu \nu} V (\bar \phi))
\end{equation}
where $g_{\mu \nu}  V (\bar \phi)$ is the energy-momentum tensor of the vacuum.
Comparing the energy-momentum tensor $T_{\mu \nu}$ with $g_{\mu \nu} V (\bar \phi)$,
the pressure exerted by the vacuum and the energy density have opposite sign,
$P_\Lambda = - \rho_\Lambda = - V (\bar \phi)$.
The vacuum energy density multiplied by $8 \pi G_N$ is usually called the cosmological constant $\Lambda$,
that is, $\Lambda = 8 \pi G_N V (\bar \phi)$.
From equation (\ref{eist}) in the Minkowski coordinates, the cosmological constant thus acts like a fluid with
the effective mass density $\rho_\Lambda = \Lambda / (8 \pi G_N)$ and the pressure
$\ P_\Lambda = - \rho_\Lambda$.

The consequence of a non-zero value for the cosmological constant has been extensively
discussed \cite{Wein0,Lind} in the context of dynamics of the universe.
The cosmological constant prevents static universe from collapsing against gravity since its positive sign plays as a universal repulsive force
for space.
In the inflation scenario \cite{Guth}, a scalar field with a non-zero vacuum energy density serves as a driving force for the exponential expansion.
This can be interpreted as that the effective cosmological constant would be a large positive constant in the early universe, which might
reduce to its present small positive value after series of phase transitions in the early universe.

The cosmological constant problem or equivalently the vacuum energy
problem is resolved in the following subsection from the gauge theory point of view.

\subsection{Resolution of the Cosmological Constant}

QG as a gauge theory exhibits reasonable explanations to the known phenomenological,
cosmological constant problem; the effective cosmological constant $\Lambda_e$ is
connected with the effective vacuum energy density $V_e (\bar \phi) = \langle \rho_m \rangle_e$
linked by the gauge boson mass $M_G$. For the gauge boson with the energy $E$ higher
than its mass $M_G$, the interaction can be overall repulsion before the phase
transition and for the gauge boson with the energy lower than its mass, the
interaction is overall attraction after the phase transition in the matter space. The
effective cosmological constant $\Lambda_e$ representing the effective vacuum energy
density decreases its value through DSSB from a false vacuum to a true vacuum, induced
by the condensation of singlet gravitons. A vacuum state is stable if vacuum
expectation values for gauge bosons are at a true minimum of the effective potential.
However, if vacuum expectation values are at a local minimum that is higher than the
true minimum, then this vacuum will be metastable. A metastable, false vacuum state
corresponding to a local minimum above the Planck energy will decay into the stable
true vacuum corresponding to the true minimum by a tunneling process. The normal
vacuum or false vacuum has several characteristics: isotropy, homogeneity, degeneracy,
etc. as much as the number of directions. The effective cosmological constant
$\Lambda_e$ represented by the graviton mass $M_G$ decreases through the DSSB
mechanism, which is triggered by the condensation of the singlet gauge field.

The extremely small deviation of the flat universe,
\begin{equation}
\Omega - 1 = 10^{-61} ,
\end{equation}
is required by quantum gauge theory, is confirmed by experiments BUMERANG-98 and MAXIMA-1
\cite{Jaff}, and is constrained by the space time quantization in the order of $10^{30}$
in one dimension. The vacuum represented by massive gauge bosons is thus quantized by
the maximum wavevector mode $N_R = i/(\Omega -1)^{1/2} \approx 10^{30}$ and the total
gauge boson number $N_G = 4 \pi N_R^3/3 \approx 10^{91}$; the effective gauge boson
mass $M_G^{e} = i k_G = (M_G^2 - E^2)^{1/2}$ is quantized by the maximum wavevector
mode $N_R$. The effective  cosmological constant is defined by $\Lambda_e = 8 \pi G_N
M_G^4$ where the gauge boson mass square is given by $M_G^2 = M_{Pl}^2 - g_f g_g^2
\langle \phi \rangle^2 = g_f g_g^2 [A_{0}^2 - \langle \phi \rangle^2]$ below the
Planck energy. The effective cosmological constants are respectively calculated from
the graviton mass $M_G \approx 10^{19}$ GeV at the Planck era and the nearly massless
gauge boson mass $M_G \approx 10^{-12}$ GeV at the present era: the effective
cosmological constants are $\Lambda_{Pl} > \Lambda_{EW} > \Lambda_{QCD} >
\Lambda_{0}$, where $\Lambda_e = \Lambda_{Pl} = M_{Pl}^2 \approx 10^{38}$
$\textup{GeV}^2$ at the Planck epoch, $\Lambda_e = \Lambda_{EW} \approx 10^{-30}$
$\textup{GeV}^2$ at the weak epoch, $\Lambda_e = \Lambda_{QCD} \approx 10^{-42}$
$\textup{GeV}^2$ at the strong epoch, and $\Lambda_e = \Lambda_0 = 3 H_0^2 \approx
10^{-84}$ $\textup{GeV}^2$ at the present epoch. This implies the reduction of zero
modes through the singlet gauge boson condensation leading to the current anomaly. The
effective vacuum energy density is connected with the effective cosmological constant
by $V_e (\bar \phi) = \langle \rho_m \rangle_e = M_G^4 = \Lambda_e/8 \pi G_N$, which
shows the important role of the gauge boson mass $M_G$: as a function of the singlet
graviton condensation $\bar \phi =\langle \phi \rangle$ parameterized by a scalar
field, the effective potential is subtracted by $V (\bar \phi) = \frac{\Lambda}{8 \pi
G_N} = - 2 M_{Pl}^2 g_f g_g^2 \langle \phi \rangle^2 + g_f^2 g_g^4 \langle \phi
\rangle^4 = \mu^2 \langle \phi \rangle^2 + \lambda \langle \phi \rangle^4$ according
to equations (\ref{gama}) and (\ref{comp}). The condensation of the singlet gauge
boson, represented by the bare cosmological constant, cancels the vacuum energy and
makes the accelerating expansion of the universe. At the Planck energy, the effective
vacuum energy density $\langle \rho_m \rangle_e$ has the energy density of $10^{76} \
\textup{GeV}^4$ but below the Planck energy, the condensation of singlet gauge fields
decreases it to the present value $10^{-47} \ \textup{GeV}^4$: $\langle \rho_m
\rangle_e \approx 10^8 \ \textup{GeV}^4$ at the weak energy and $\langle \rho_m
\rangle_e \approx 10^{-4} \ \textup{GeV}^4$ at the QCD energy. Since $\bar \phi
\approx 0$ above the Planck scale but $\bar \phi \approx 10^{19}$ GeV at the present
universe scale, the discrepancy in the vacuum energy density leading to $10^{122} \
\textup{GeV}^4$ is consistent with the theoretical large value of the effective vacuum
energy density at the Planck scale and the experimental small value at the present
universe scale. It is realized that the total conserved particle number of gauge
bosons is $10^{91}$, the energy per gauge boson is $10^{19}$ GeV at the Planck epoch,
and the energy per gauge boson is $10^{-12}$ GeV at the present epoch; the static
maximum (minimum) gauge boson energy is $10^{19}$ GeV ($10^{-12}$ GeV) at the Planck
epoch and is $10^{-12}$ GeV ($10^{-42}$ GeV) at the present epoch.

The gauge boson number density is given by $n_G = M_G^3$: $n_{Pl} \approx 10^{57} \
\textup{GeV}^3 \approx 10^{98} \ \textup{cm}^{-3}$ at the Planck scale, $n_{EW}
\approx 10^{6} \ \textup{GeV}^3 \approx 10^{47} \ \textup{cm}^{-3}$ at the weak scale,
$n_{QCD} \approx 10^{-2} \ \textup{GeV}^3 \approx 10^{39} \ \textup{cm}^{-3}$ at the
strong scale, and $n_{0} \approx 10^{-36} \ \textup{GeV}^3 \approx 10^{5} \
\textup{cm}^{-3}$ at the present scale. Under the constraint of the extremely flat
universe, the relation $\Omega - 1 = (\langle \rho_m \rangle - \Theta \rho_m)/\rho_G
-1 = - 10^{-61}$ leads to the $\Theta$ constant $\Theta = 10^{-61} \ \rho_G/\rho_m$,
which will be further discussed. If the matter density in the universe is $\rho_m
\simeq \rho_c \simeq 10^{-47} \ \textup{GeV}^4$ and is conserved, the $\Theta$
constant depends on the gauge boson mass $M_G$: $\Theta = 10^{-61} \ M_G^4/\rho_c$.
$\Theta$ values becomes $\Theta_{Pl} \approx 10^{61}$, $\Theta_{EW} \approx 10^{-4}$,
$\Theta_{QCD} \approx 10^{-10}$, and $\Theta_{0} \approx 10^{-61}$ at different
stages. This is consistent with the observed results, $\Theta < 10^{-9}$ in the
electric dipole moment of the neutron \cite{Alta} and $\Theta \simeq 10^{-3}$ in the
neutral kaon decay \cite{Chri} as CP violation parameters.

The effective cosmological constant corresponds to the gauge boson mass, which is
large in the early small system before phase transition but is small in the later
large system after phase transition: the nearly zero cosmological constant represents
the nearly massless gauge boson responsible for the universe expansion discussed in
the following section. The DSSB of gauge symmetry and discrete symmetries might give
rise to the current anomaly like the axial current anomaly in QCD \cite{Hoof2}.
The true vacuum as the physical vacuum is achieved from the normal vacuum, which
possesses larger symmetry group than the physical vacuum, through DSSB. The instanton
mechanism as the vacuum tunneling is expected in the Euclidean spacetime. In the
universe evolution, the breaking of discrete symmetries, P, C, CP, and T, is expected
due to the condensation of the singlet gauge boson and is closely relevant for the
baryon asymmetry. The $\Theta$ vacuum term in (\ref{thet}) explicitly violates CP
symmetry: $\Theta = 10^{-61} \rho_G/\rho_m$ in equation (\ref{thev}). According to the
observation for the electric dipole moment of the neutron $d_n = 2.7 \approx 5.2
\times 10^{-16} \Theta \ e \ \textup{cm}$, $\Theta \leq 10^{-9}$ at the strong scale \cite{Alta}.
CP violation due to the $\Theta$ vacuum term makes the baryon asymmetry $\delta_B
\approx 10^{-10}$ observed at present \cite{Stei0}. According to the baryon asymmetry, the fermion
matter current $J_\mu$ is conserved at the Planck epoch but the antifermion current
$\bar J_\mu$ is not conserved, that is, $\partial_\mu J_\mu = 0$ but $\partial_\mu
\bar J_\mu = \frac{g_g^2}{16 \pi^2} Tr G^{\mu \nu} \tilde G_{\mu \nu}$: this makes the
fermion asymmetry $\delta_f = \frac{N_f}{N_{t \gamma}} = \frac{10^{91}}{10^{88}} =
10^3$ with the fermion mass $10^{-12}$ GeV as discussed in the following section. The
bare vacuum energy density $\langle \rho_m \rangle = 10^{76} \ \textup{GeV}^4$ at the Planck scale
is relevant for the $\Theta$ vacuum, which represents the surface term because it is
total derivative and the effective cosmological constant decreases as the system
expands. The vacuum energy density does not affect the perturbative aspect but does
affect the nonperturbative aspect of the system. The massive gauge boson with the
Planck mass is transformed to the nearly massless gauge boson with the mass $10^{-12}$
GeV at present. Massless gauge bosons (photons) respectively appear during phase
transition at the Planck epoch and at the present epoch: $E_\gamma \approx 10^{19}$
GeV and $E_\gamma \approx 3 \times 10^{-13}$ GeV respectively. DSSB due to the
initially large cosmological constant $\Lambda_e$ derives the inflation of the system
in the order of $10^{30}$, which make the system expand exponentially \cite{Guth}.
This is totally compatible with the observed expectation of the small effective
cosmological constant $10^{-84} \ \textup{GeV}^2$ or energy density $10^{-47} \
\textup{GeV}^4$ at the very low energy scale of the present universe and with the
theoretical expectation of the large effective cosmological constant or energy density
$10^{76} \ \textup{GeV}^4$ at the Planck scale of the early universe. This scheme may
accordingly resolve the cosmological constant problem and may be compatible with observations
BUMERANG-98 and MAXIMA-1 \cite{Jaff}.

\subsection{Modification of General Relativity: $\Theta$ Constant and $\Lambda$ Constant}

The value of $\Theta$ also makes the relation of the matter energy density $\rho_m$ to
the effective cosmological constant $\Lambda_e$ because of the relation $\Lambda_e = 8 \pi G_N M_G^4$:
\begin{equation}
\rho_m = \frac{\rho_G}{10^{61} \Theta} = \frac{M_G^4}{10^{61} \Theta} = \frac{\Lambda_e}{10^{61} \Theta 8 \pi G_N}
\end{equation}
where $\rho_m = T_{00}$ with the energy-momentum tensor $T_{\mu \nu}$.
One possible extension of Einstein's field equation (\ref{eisc}) is obtained by using the above relation:
\begin{equation}
\label{eisc1} R_{\mu \nu} - \frac{1}{2} g_{\mu \nu} R = - 8 \pi G_N \tilde T_{\mu \nu}
= - 8 \pi G_N (T_{\mu \nu} - g_{\mu \nu} 10^{61} \Theta \rho_m)
\end{equation}
or, under the assumption of relativistic matter particle $T/2 = \rho_m$,
\begin{equation}
\label{eisc5}
R_{\mu \nu} - \frac{1}{2} g_{\mu \nu} R = - 8 \pi G_N \tilde T_{\mu \nu} = - 8 \pi G_N (T_{\mu \nu} - \frac{10^{61} \Theta}{2} g_{\mu \nu} T)
\end{equation}
where the first term of the right hand side stands for matter energy and the second term stands for vacuum energy.
Since the vacuum energy density is $V (\bar \phi) = \rho_G = 10^{61} \Theta \rho_m$,
(\ref{eisc1}) is the equivalent form with (\ref{eist}).
The $\Theta$ constant thus plays the role relating two different worlds, the matter world and the vacuum world.
Using $\Theta$ values, effective cosmological constants are obtained:
$\Lambda_e = 8 \pi G_N M_G^4 = 10^{38} \ \textup{GeV}^2$ at the Planck epoch,
$\Lambda_e = \Lambda_{EW} \approx 10^{-30}$ $\textup{GeV}^2$ at the weak epoch,
$\Lambda_e = \Lambda_{QCD} \approx 10^{-42}$ $\textup{GeV}^2$ at the strong epoch, and
$\Lambda_e = \Lambda_0 = 3 H_0^2 \approx 10^{-84}$ $\textup{GeV}^2$ at the present epoch.
At the present universe with $\Theta_0 = 10^{-61}$, the modified general relativity (\ref{eisc5}) becomes
\begin{equation}
\label{eisc6}
R_{\mu \nu} - \frac{1}{2} g_{\mu \nu} R = - 8 \pi G_N (T_{\mu \nu} - \frac{1}{2} g_{\mu \nu} T)
\end{equation}
which is identical with general relativity (\ref{eist}) since $T/2 = \rho_m = \Lambda_0/8 \pi G_N \simeq \rho_c$.
Note that the left hand side and the right hand side have completely symmetric forms in two tensors $R_{\mu \nu}$ and $T_{\mu \nu}$.

\section{Classical Tests}

In the previous sections, Newton gravitation constant is defined
as the effective coupling constant and the effective cosmological
constant is connected to the effective vacuum energy due to
massive gauge bosons.
In this section, only classical tests for QG are concentrated while
quantum tests for QG are separately discussed in subsequent papaer \cite{Roh11}.

Whatever the gauge group $G$ for QG is, a gauge theory for gravity
can exhibit classical tests verified by Einstein's general
relativity in the weak field limit \cite{Wein}. Based on QG, the
behavior of the gauge boson depends on its energy compared with
its mass. If the gauge boson has energy $E > M_{Pl}$, the
gravitational propagation becomes like $e^{ik(r - l_{Pl})}$. If $E
< M_{Pl}$, propagation  $e^{-M_G^{e} (r - l_{Pl})}$ is limited
within the Planck scale $l_{Pl} \approx 1/M_{Pl}$ scale; this is
the Yukawa interaction with the massive graviton. When $E
\rightarrow M_{Pl}$, the massive gauge boson mediates the constant
gravitational potential. When $E < M_{Pl}$, the massive gauge
boson mediates the Yukawa type potential $e^{-M_G^{e} (r -
l_{Pl})}$ where quadrupole moment interactions lead to Einstein's
general relativity. Then the Newtonian approximation limit of
general relativity is the familiar Newtonian mechanics in the
macroscopic scale. The gauge boson with the energy less than the
Planck mass provides Newtonian mechanics outside the Planck scale;
the relativistic correction to the rest mass is the direct reason
for the modification of Newtonian mechanics as the quadrupole
moment interaction. Newtonian mechanics in the solar system is the
case of the large baryon number $(B = N_B >> 1)$:
in the relation between the
gauge boson and fermion mass $M_G = m g_f g_g^2 \sqrt{N_{sd}}$,
$N_{sd} << 1$ in the case of $m >> M_G$.

\subsection{Relations among Quantum Gauge Theory, General Relativity, and Newtonian Mechanics}

QG based on quantum mechanics, gauge invariance, and special relativity is connected with general relativity
based on equivalence principle and general coordinate invariance through the parameter $\Theta$
and with Newtonian mechanics in terms of the multipole expansion of the Yukawa gravitational potential.
The brief outline is given as follows.

\subsubsection{Relation between Quantum Gauge Theory and General Relativity}

According to QG, the $\Theta$ vacuum generates matter as the result of the violation of discrete symmetries during DSSB:
the parameter $\Theta$ provides the relation between the vacuum and matter as indicated by the baryon asymmetry.
The explicit relation between QG and Einstein's generality is given by
\begin{equation}
\label{eisc2} R_{\mu \nu} - \frac{1}{2} g_{\mu \nu} R = - 8 \pi G_N \tilde T_{\mu \nu}
= - 8 \pi G_N (T_{\mu \nu} - \frac{10^{61} \Theta}{2} g_{\mu \nu} T) .
\end{equation}
The weak field limit of general relativity can be, apart from the vacuum term $\Lambda = \Theta 10^{61} 8 \pi G_N T/2$,
obtained by decomposing $g_{\mu \nu} = \eta_{\mu \nu} + h_{\mu \nu}$ with $h_{\mu \nu} << 1$:
\begin{eqnarray}
\label{eisc3}
\Box h_{km} & = & -16 \pi G_N (T_{km} - \eta_{km} T/2), \nonumber \\
\Box (h_j^i & - & \frac{1}{2} \delta^i_j h) = -16 \pi G_N T_j^i
\end{eqnarray}
where the first wave equation describes gravitational radiation as well as the
response of the gravitational field to the source $T_{\mu \nu}$. In the weak field
limit of Einstein's generality relativity, the matter interaction is postulated as the
quadrupole (tensor) interaction with the angular momentum $l=2$ (or $s=2$), which possesses
complete symmetric configuration. The energy is the source in general relativity,
which mediates only attractive force. This information is consistent with the
interpretation of QG since the gravitational electric quadrupole moment must be of
even parity for matter and be of attraction: the odd parity for gravitational magnetic
quadrupole moment disappears so as to conserve parity operation for matter. Massive
gauge bosons with an $s=2$ multiplet are $A_4 \sim A_8$ if QG is an $SU(3)$ gauge
theory. A massive gauge boson with $s =2$ has the mass $M_G \approx 10^{19}$ GeV at
the Planck scale and the mass $E_\gamma \approx 10^{-12}$ GeV at the present scale.
The massless gauge boson with $l=2$ (or $s=2$) in general relativity, conventionally
regarded as the graviton, is realized as the massless photon with $l=2$ (or $s=2$) in QG in this aspect.
The massless gauge boson with $s=1$ or $s=2$, known as the photon, has the energy $E_\gamma \approx 10^{18}$ GeV at the Planck
scale and the energy $E_\gamma \approx 10^{-13}$ GeV as CMBR at the present scale.

\subsubsection{Relation between General Relativity and Newtonian Mechanics}

Newtonian mechanics is known as the weak field approximation or
the Newtonian approximation of general relativity. Newtonian
approximation is specified by the following assumptions. The
motion of particles are nonrelativistic ($ v << c$), the
gravitational fields are weak in the sense that $g_{ij} =
\eta_{ij} + h_{ij}$ with $h_{ij} << 1$, and the fields change
slowly with the time. The inequality suggests that powers of
$h_{ij}$ higher than $2$ in the action principle and higher than
$1$ in the field equations are neglected and time derivatives are
ignored in comparison with space derivatives. In the quadrupole (tensor)
interaction in the weak field limit, the mass of particle is the
source, the gravitational potential $\phi = h_{00}/2$ is the
Coulomb type potential ($\sim 1/r$), and there is only attractive
force: $\nabla^2 \phi = - 4 \pi G_N \rho_m$ or $\phi = - G_N m/r$
is derived from (\ref{eisc3}).

\subsubsection{Relation between Quantum Gauge Theory and Newtonian Mechanics}

The spherical mass is roughly regarded as the gravitational
intrinsic electric monopole. The generation of the matter mass $m$
is closely related to the spontaneous breaking of gauge and chiral
symmetries. As the temperature of the system decreases, the gauge
boson mass decreases due to the condensation of the singlet gauge
boson and the matter particle mass $m$ increases with the decrease
of the difference number $N_{sd}$. It is realized that
the Yukawa type gravitational potential $\phi (r) = e^{- M_G^e (r
- l_{Pl})}$ is applied when $M_G^{e2} = (M_{Pl}^2 - E^2) > 0$ but
the plain wave propagation $\phi (r) = e^{ik (r - l_{Pl})}$ is
applied when $k^2 = (E^2 - M_{Pl}^2) > 0$.

Newtonian mechanics is the Yukawa gravitational potential associated with massive gauge boson:
$M_G^e \simeq M_G \simeq M_{Pl}$ at the present universe.
Using its multipole expansion around the boundary $l_{Pl} = 1/M_{Pl}$, the gravitational potential
\begin{equation}
\label{yupo}
\phi (r) = e^{-M_G (r - l_{Pl})}
\end{equation}
becomes
\begin{equation}
\label{yupo1}
\phi (r) =  - \frac{1}{M_G r} +  \frac{2}{M_G^2 r^2} \cdot \cdot \cdot
\end{equation}
as the regular solution for multipole terms outside the Planck scale $r > l_{Pl}$ and becomes
\begin{eqnarray}
\label{yupo2}
\phi (r) = 1 - M_G r + M_G^2 r^2/2 \cdot \cdot \cdot
\end{eqnarray}
as the regular solution for multipole terms inside the Planck
scale $r < l_{Pl}$. The quadrupole expansion ($l=2$ or $s=2$) in
(\ref{yupo1}) is related to the Newtonian potential outside the
Planck scale $l_{Pl}$. Newton gravitational constant $G_N$ is the
effective coupling constant specified by the mass of the gauge
boson $M_G$: $G_N = \sqrt{2} g_f g_g^2/8 M_G^2$. It is assumed
that the interaction distance between constituent particles in
analogy with the Bohr radius is expressed by $r_i = 1/m_i g_f
\alpha_g$ with the constituent particle mass $m_i$ and the
macroscopic distance is expressed by $r = 1/m g_f \alpha_g \simeq
2 \sqrt{2} /\pi \sum_i r_i$ with the total mass given by the sum
of constituent masses $m = \sum_i m_i$. Then the collective
quadrupole term is recovered as the Newtonian gravitational
potential
\begin{equation}
\label{nupo3}
\phi (r) = - G_N m /r
\end{equation}
and the gravitational potential energy
\begin{equation}
\label{nupo4}
U (r) = - G_N m_a m_b /r
\end{equation}
where $m_a$ and $m_b$ are the masses of particles. Note that the
quadrupole term of the potential (\ref{yumur}) leads to the
gravitational potential energy (\ref{nupo4}) outside the Planck
scale. The matter mass $m$ is connected with the gauge boson mass
by $M_G = \sqrt{\pi} m g_f \alpha_g \sqrt{N_{sd}}$ with the difference
number $N_{sd}$ in intrinsic two-space dimensions. It is
realized that since the force in Newtonian mechanics is $F = - G_N
m_a m_b/r^2 \sim - 1/N_{sd} r^2$, the great number of gravitons with
the mass $M_{Pl}$ are, roughly $1/N_{sd} \approx 10^{70}$ in the
solar system, involved in gravitational interactions because, for
example, the sun has the mass $m_S \sim 10^{57}$ GeV and the earth
has the mass $m_E \sim 10^{51}$ GeV. Newtonian mechanics is the
residual interaction of the graviton exchange just as the nuclear
interaction is the residual interaction of the gluon exchange.
This suggests that gravitational dipoles known as the Cooper pairs are
formed and they do the quadrupole (tensor) interaction, which is
approximated by the Coulomb type potential outside the Planck
scale in the special case of the $z \approx 0$ plane.

\subsection{Classical Test of Quantum Gravity}

QG may be related to general relativity in the weak field limit and Newton mechanics
in terms of gauge bosons with the energy less than the Planck mass.
The energy per mass derived from QG is connected with Newtonian mechanics and its relativistic correction so that QG may also derive
classical tests verified by general relativity in the weak field limit.

\subsubsection{Newtonian Mechanics and Relativistic Correction}

The collective quadrupole (tensor) interaction with $l=2$ (or
$s=2$) is the matter interaction: this is the gravitational
electric quadrupole to satisfy the parity operation of the matter
space. Gravitational magnetic dipole is possible but gravitational
electric dipole is not possible due to parity condition.
Gravitational magnetic monopole violates parity operation and this
might be the reason for no detection of magnetic monopole in the
universe as indicated by the dual Meissner effect.

Based on the dual Meissner effect in gravitation, the graviton with the energy lower than the Planck mass
can not propagate outside the Planck scale and the gravitational electric field is excluded in the gravitational magnetic superconducting state.
When matter particles interact by the exchange of gravitons,
their quadrupole interactions hold even parity outside the Planck scale.
The small correction to Newtonian mechanics comes from the special relativistic effect to the rest matter mass $m$, which becomes
the effective mass $m/\sqrt{1 - u^2}$ with the velocity $\vec u$:
$m \rightarrow m/(1 - u^2/2) \simeq m (1 + u^2/2)$ if $u << 1$.
The gravitational potential energy in (\ref{nupo4}) leads to
\begin{equation}
\label{nupo5}
U (r) = - G_N \frac{m_a m_b} {r} (1 + u^2)
\end{equation}
Note that the longitudinal component of $\vec u$ gives repulsion and the
transverse component of $\vec u$ gives attraction.

To apply the above description to the solar system, assume that an object with the mass $m_a = m$ orbits around a very massive object with $m_b = M$.
The energy per mass including the kinetic energy and potential energy can be written by
\begin{eqnarray}
\label{qgsu4}
\varepsilon & = & \frac{1}{2} (\dot{r}^2 + r^2 \dot{\varphi}^2 ) - r^2\dot{\varphi}^2 \frac{G_N M}{r} - \frac{G_N M}{r} \nonumber \\
& = & \frac{1}{2} (\dot{r}^2 + r^2 \dot{\varphi}^2 [1 - 2 \frac{G_N M}{r}]) - \frac{G_N M}{r}
\end{eqnarray}
where the first two terms are the kinetic energies of the mass $m$ in the
polar coordinates and the last two terms are the Newtonian potential and its relativistic correction.
Note that $\vec u$ have only the transverse components and
the transverse particle velocity $u_\perp =  r \dot{\varphi}$ is used in the above.
In general relativity, total energy per mass
is exactly identical to equation (\ref{qgsu4}).

This verifies that gravity may be described by a gauge theory since both general
relativity in the weak field limit and gauge theory with the DSSB mechanism can
produce the identical energy per mass. The effective spin $2$ interaction in general
relativity is verified as the intrinsic or extrinsic quadrupole interaction ($s=2$ or
$l=2$) with the effective coupling constant $G_N$ in gauge theory. Equation
(\ref{qgsu4}) may provide explanations to the typical classical tests, the precession
of the perihelion and the deflection of light \cite{Berr}.

\section{Comparison between Quantum Gravity and General Relativity}

This section is devoted to summarize and to convince QG as an $SU(N)$ gauge theory
beyond general relativity: classical tests for QG are discussed in the previous
section and quantum tests for QG are described in subsequent papaer \cite{Roh11}.

The potential QG as a gauge theory may resolve serious problems of GUTs and the standard
model: different gauge groups, Higgs particles, the inclusion of gravity, the proton
lifetime, the baryon asymmetry, the family symmetry of elementary particles,
inflation, fermion mass generation, etc. This scheme may also provide possible
resolutions to the problems of Einstein's general relativity or the standard hot big
bang theory: the spacetime singularity, cosmological constant, quantization, baryon
asymmetry, structure formation, dark matter, flatness of the universe, and
renormalizability, etc. Furthermore, the typical predictions of QG are compatible with
recent experiments, BUMERANG-98 and MAXIMA-1 \cite{Jaff}: the flat universe,
inflation, vacuum energy, dark matter, repulsive force, CMBR, etc.

Comparison between QG and general relativity is summarized by
Table \ref{cmqg}. QG is quantum theory holding principles of gauge
invariance and special relativity while Einstein's general theory
of relativity is classical theory holding principles of
equivalence and general relativity.  Newly introduced concepts in
QG are the flat universe rather than the curved universe,
dynamical spontaneous symmetry breaking (DSSB) rather than
spontaneous symmetry breaking, gauge group hierarchy $G \supset H
\supset SU(2)_L \times U(1)_Y \times SU(3)_C$, coupling constant hierarchy
$(\alpha, \alpha/3, \alpha/4, \alpha/12, \alpha/16)$ for both weak
and strong interactions, effective coupling constant hierarchy
$G_N \supset G_F \supset G_R$, effective cosmological constant
$\Lambda_e = 8 \pi G_N M_G^4$, massive gauge boson mass square
$M_G^2 = M_{Pl}^2 - g_f g_g^2 \langle \phi \rangle^2 = g_f g_g^2
[A^2_0 - \langle \phi \rangle^2]$, massless gauge bosons as
Nambu-Goldstone bosons including cosmic microwave background radiation \cite{Penz},
the dynamical spontaneous
breaking of discrete symmetries (C, P, T, and CP), Newtonian
mechanics as quadrupole (tensor) interactions, the possible
modification of Einstein's general relativity, etc.
Quantum tests will be discussed in subsequent paper \cite{Roh11} are
strongly interacting massive particles in addition to strongly interacting massive particles
as the candidate of dark
matter, the universe inflation with the order of magnitude
$10^{30}$, the baryon asymmetry $\delta_B \approx 10^{-10}$, the
relation between time and gauge boson mass, the internal and
external quantization of spacetime, proton decay time much longer
than one predicted by GUTs, mass generation mechanism with the
surface effect, etc. The gauge boson mass $M_G \simeq 10^{-12}$
GeV associated with the cosmological constant $\Lambda_0 \simeq
10^{-84} \ \textup{GeV}^2$ and the Hubble constant $H_0 \simeq
10^{-42}$ GeV especially indicates gauge theories for new
fundamental forces responsible for the universe expansion and CMBR.
The repulsive force at the present universe is recently suggested
in BUMERANG-98 and MAXIMA-1 experiments \cite{Jaff}.

\section{Conclusions}

This study toward quantum gravity (QG) proposes an $SU(N)$ gauge
theory with the $\Theta$ vacuum term as a trial theory, which
suggests that a certain group $G$ for gravitational interactions
leads to a group $SU(2)_L \times U(1)_Y \times SU(3)_C$ for weak and strong
interactions through dynamical spontaneous symmetry breaking
(DSSB) leading to a current anomaly; the group chain is $G \supset
SU(2)_L \times U(1)_Y \times SU(3)_C$. DSSB consists of two simultaneous
mechanisms; the first mechanism is the explicit symmetry breaking
of gauge symmetry, which is represented by the gravitational
factor $g_f$ and the gravitational coupling constant $g_g$, and
the second mechanism is the spontaneous symmetry breaking of gauge
fields, which is represented by the condensation of singlet gauge
fields. Newton gravitation constant $G_N$ originates from the
effective coupling constant for massive gravitons,
$\frac{G_N}{\sqrt{2}} = \frac{g_f g_g^2}{8 M_G^2}$ with $M_G =
M_{Pl} \approx 10^{19}$ GeV: the effective coupling constant chain
is $G_N \supset G_F \times G_R$ for gravitation, weak, and strong
interactions respectively. This scheme relates the effective
cosmological constant to the effective vacuum energy associated
with massive gauge bosons, $M_G^2 = M_{Pl}^2 - g_f g_g^2
\langle \phi \rangle^2$, and provides a plausible explanation for both the
present small and the early large value of the cosmological
constant; the condensation of the singlet gauge field $\langle \phi \rangle$
induces the current anomaly and subtracts the gauge boson mass as
the system expands. This proposal thus suggests a viable solution
toward such longstanding problems as the quantization of gravity
and the cosmological constant. The crucial point is that DSSB
mechanism is adopted to all the interactions characterized by
gauge invariance, physical vacuum problem, and discrete symmetry
breaking. In this approach, DSSB is currently underway in the
evolution of the flat universe and Newton gravitation coupling
constant $G_N$ represents the massive gauge boson propagation just
as Fermi electroweak coupling constant $G_F$ does. Newton
gravitation constant has the dimension of inverse energy square
due to massive gravitons. A gauge theory allows to have the masses
of gauge bosons without spoiling the renormalizability; the
renormalizability of a gauge theory with DSSB was demonstrated by
't Hooft. The dimensionality of the effective gravitation coupling
constant, which prevents gravity from the renormalizability, may
be removed in terms of the concept of DSSB. The compatibility
between gravitation theory and gauge theory may be extended to the
quantization and renormalization of gravity and the
non-renormalizability problem of gravity may be resolved in the
potential gauge theory for gravity.

This study claims that the effective cosmological constant defined by $\Lambda_e = 8
\pi G_N M_G^4$ is related to the effective vacuum energy $V_e (\bar \phi) = \langle \rho_m \rangle_e
= M_G^4$ represented by the gauge boson mass $M_G$, which plays an important role
specially at the very early and later stage of gravitational evolution: $\Lambda_e
\approx 10^{38} \ \textup{GeV}^2$ or $\langle \rho_m \rangle_e \approx 10^{76} \ \textup{GeV}^4$ at
the Planck epoch and $\Lambda_e = \Lambda_0 \approx 10^{-84} \ \textup{GeV}^2$ or
$\langle \rho_m \rangle_e = \langle \rho_m \rangle_0 \approx 10^{-47} \ \textup{GeV}^4$ at the present epoch. This
proposal may resolve the cosmological constant problem by interpreting that the
condensation of singlet gauge bosons exactly cancels the bare vacuum energy density
$\langle \rho_m \rangle$ due to the condensation of fermions through the universe expansion. The
effective vacuum energy linked by the gauge boson mass becomes significant during
phase transition in the early universe and makes the universe expand exponentially.
This scheme is therefore
consistent with both the experimental expectation of the very small effective
cosmological constant at the present epoch and the theoretical expectation of the very
large cosmological constant at the early universe; the condensation of singlet gauge
bosons makes gauge bosons lighter as the universe expands and phase transition takes
place. The $\Theta$ constant is parameterized by $\Theta = 10^{-61} \ \rho_G/\rho_m =
10^{-61} \ \Lambda_e/8 \pi G_N \rho_m$ with the vacuum energy density $\rho_G = M_G^4$
and the matter energy density $\rho_m \simeq \rho_c \simeq 10^{-47} \ \textup{GeV}^4$.
This suggests the possible modification of general relativity in terms of the $\Theta$
constant rather than the cosmological constant $\Lambda$:
\begin{math}
R_{\mu \nu} - \frac{1}{2} g_{\mu \nu} R
= - 8 \pi G_N (T_{\mu \nu} - \frac{10^{61} \Theta}{2} g_{\mu \nu} T) .
\end{math}

A QG is consistent with general relativity in the weak field limit as well as
Newtonian mechanics since classical tests can be successfully shown from the gauge
theory point of view. The relation between QG and Einstein's general relativity is
explicitly given in terms of cosmological parameters like $\Lambda$ and $\Theta$. The
graviton with the energy lower than the Planck mass has the Yukawa type potential
$\phi(r) = e^{- M_G^e (r - l_{Pl})}$ with the effective gauge boson mass $M_G^{e}= i
k_G = (M_G^2 - E^2)^{1/2}$ while the graviton with the energy equal to the Planck mass
has the Coulomb potential energy $-G_N m^2/r$. This implies the effective interaction
of the graviton below the Planck energy is proportional to $G_N J_\mu J^{\mu \dagger}$
in the low momentum transfer. When $E < M_{Pl}$, the massive gauge boson mediates the
Yukawa type potential $\phi (r) = e^{-M_G^{e} (r - l_{Pl})}$ where the quadrupole
moment term leads to the weak field approximation of Einstein's general relativity:
$\phi(r) = h_{00}/2$ with $g_{ij} = \eta_{ij} + h_{ij}$ and $h_{ij} <<1$. Newtonian
mechanics, which is the Newtonian approximation of the weak field limit of general
relativity, is interpreted as collective quadrupole (tensor) interactions of gravitons with the
energy less than the Planck mass ($E < M_{Pl}$). The energy per mass obtained from
this scheme, which has the Newtonian potential for the rest mass and the relativistic
correction to the rest mass, is the same with one obtained from general relativity;
the energy per mass can leads to the classical tests verified by general relativity.
This work might accordingly give rise to a turning point toward the unification of
fundamental forces in nature. QG as a potential gauge theory may resolve the
longstanding problems, the quantization of gravity and the cosmological constant
problem. This proposal beyond the standard model thus implies that a potential,
ultimate theory of nature may be found in the form of quantum field theory.

Notable accomplishments of this work are summarized as follows. A potential QG as a
gauge theory is introduced and its relation to Einstein's general relativity is
suggested. The origin of Newton gravitation constant $G_N$ is illustrated as the
Yukawa coupling of the massive graviton. The origin of the cosmological constant is
given as the vacuum energy density represented by the gauge boson mass; the total
gauge boson number $10^{91}$ in the universe is predicted as a conserved good quantum
number. The DSSB of
local gauge symmetry and global chiral symmetry triggers the baryon current anomaly.
Newton gravitation constant $G_N$ as the effective coupling constant provides the
construction of the dimensionless coupling constant through the condensation of
singlet gauge fields, which makes this theory renormalizable.
The classical tests can be shown as the illustration of QG as a gauge theory even
though the exact group for QG is not uncovered yet.  This
scheme may also provide possible resolutions to the problems of Einstein's general
relativity or the standard hot big bang theory: the spacetime singularity,
cosmological constant, quantization, baryon asymmetry, structure formation, dark
matter, flatness of the universe, and renormalizability, etc. This work significantly
contributes to the unification of fundamental forces from the viewpoint of the gauge
field theories since all the known forces might be formulated in terms of gauge
theories; the standard hot big bang theory with many pros and cons \cite{Peeb,Arp} may
be replaced by this scheme toward the theory of everything.

\onecolumn

\begin{table}
\caption{\label{fefu} Features of Fundamental Interactions}
\end{table}
\centerline{
\begin{tabular}{|c|c|c|c|c|} \hline
Feature & Gravity & Electromagnetic & Weak & Strong \\ \hline \hline
Gauge Boson & Graviton & Photon & Intermediate Boson & Gluon \\ \hline
Source & Spin (?) & Electric Charge & Isospin & Color \\ \hline
Coupling Constant & $\alpha_g (M_{Pl}) \simeq 0.02$ (?) & $\alpha_e \simeq 1/137$ &
$\alpha_i (M_Z) \simeq 0.12$ & $\alpha_s (\Lambda_{QCD}) \simeq 0.48$ \\ \hline
Gauge Boson Mass & $10^{19}$ GeV & 0 & $10^2$ GeV & $10^{-1}$ GeV \\ \hline
Effective Coupling & $G_N \simeq 10^{-38} \ \textup{GeV}^{-2}$ & &
$G_F \simeq 10^{-5} \ \textup{GeV}^{-2}$ & $G_R \simeq 10^{1} \ \textup{GeV}^{-2}$ \\ \hline
Cross Section & $10^{-70} \ \textup{m}^2$ & $10^{-33} \ \textup{m}^2$ &
$10^{-44} \ \textup{m}^2$ & $10^{-30} \ \textup{m}^2$ \\ \hline
Lifetime & $10^{57}$ s & $10^{-20}$ s & $10^{-8}$ s & $10^{-23}$ s \\ \hline
\end{tabular}
}

\vspace{1cm}

\begin{table}
\caption{\label{cmqg} Comparison between Quantum Gravity and
General Relativity}
\end{table}
\centerline{
\begin{tabular}{|c|c|c|} \hline
Classification & QG & General Relativity \\ \hline \hline Exchange
Particles & massive gravitons & massless gravitons \\ \hline DSSB
& yes & no \\ \hline Discrete symmetries (P, C, T, CP) & breaking
& no \\ \hline Monopole Confinement & yes & unknown
\\ \hline Cosmological constant & $\Lambda_e = 8 \pi G_N M_G^4$ &
$\Lambda_0 = 10^{-84} \ \textup{GeV}^2$ \\ \hline Inflation &
$10^{30}$ & no \\ \hline Matter mass generation & $\rho_m =
10^{-61} \rho_G \Theta$ & unknown \\ \hline Baryogenesis &
$10^{78}$ for $0.94$ GeV proton & no \\ \hline Leptogenesis &
$10^{81}$ for $0.5$ MeV electron & no \\ \hline Proton decay time
& $10^{57}$ s & unknown \\ \hline Singularity & no & yes \\ \hline
Universe & flat ($\Omega = 1 - 10^{-61}$) & curved \\ \hline Dark
matter & yes & unknown \\ \hline Interactions & monopole or dipole
& quadrupole (tensor) \\ \hline Renormalization & yes & no \\
\hline Relativity & special & general \\ \hline Principle & gauge
invariance & equivalence \\ \hline Free parameter & coupling
constant ($\alpha_g$) & effective coupling constant ($G_N$) \\
\hline
\end{tabular}
}


\begin{references}
\bibitem{Eins} A. Einstein, Die Feldgleichungen der Gravitation (The Gravitational Field Equations),
Preuss Akad. Wiss. Berlin, Sitz. Ber., pp 844 - 847 (1915);
A. Einstein, Sitz. Preuss. Akad. Wiss., 142 (1917).
\bibitem{Jaff} A. H. Jaffe et al., astro-ph/0007333;
P. de Bernardis et al., Nature {\bf 404}, 955 (2000);
S. Hanany et al., astro-ph/0005123;
A. Balbi et al., astro-ph/0005124.
\bibitem{Roh11} H. Roh, ``Toward Quantum Gravity II: Quantum Tests,'' (Submitted to Nuclear Physics B).
\bibitem{Glas} S. L. Glashow, Nucl. Phys. {\bf 22}, 579 (1961);
S. Weinberg, Phys. Rev. Lett. {\bf 19}, 1264 (1967);
A. Salam, Elementary Particle Theory, N. Svaratholm Ed., Almquist and Wiksells (1968).
\bibitem{Frit} H. Fritzsch and M. Gell-Mann, in Proceedings of the Sixteenth International Conference on High Energy Physics, Vol. {\bf 2}, Chicago (1972);
H. Fritzsch, M. Gell-Mann and Leutwyler, Phys. Lett. {\bf  B 47}, 365 (1973); S.
Weinberg, Phys. Rev. Lett. {\bf 31}, 494 (1973).
\bibitem{Geor} H. Georgi and S. L. Glashow, Phys. Rev. Lett. {\bf 32}, 438 (1974).
\bibitem{Eins1} A. Einstein, Kosmologische Betrachtungen Zur Allgemeinen Relativitatstheorie Preuss. Akad. Wiss. Sitz. Ber. (1917).
\bibitem{Wein0} S. Weinberg, Rev. of Mod. Phys., Vol. {\bf 61}, No. 1, 1 (1989) and references therein.
\bibitem{Wein} See, e.g., S. Weinberg, {\em Gravitation and Cosmology}, pp 251 - 260, Wiley (1972);
P. J. E. Peebles, {\em Principles of Physical Cosmology}, pp 265 -275, Princeton University Press (1993);
J. V. Narlikar, {\em Introduction to Cosmology}, 2nd Ed., pp 105 - 200, Cambridge (1993).
\bibitem{Adle} S. L. Adler, Phys. Rev. {\bf 177}, 2426 (1969);
J. S. Bell and R. Jackiw, Nuovo Cimento {\bf A 60}, 47 (1969).
\bibitem{Hoof2} G. 't Hooft, Phys. Rev. Lett. {\bf 37}, 8 (1976); Phys. Rev. {\bf D 14}, 3432 (1976);
R. Jackiw and C. Rebbi, Phys. Rev. Lett. {\bf 37}, 172 (1976);
C. G. Callan, R. Dashen, and D. Gross, Phys. Lett. {\bf 63 B}, 334 (1976).
\bibitem{Higg} P W. Higgs, Phys. Rev. Lett. {\bf 13}, 508 (1964).
\bibitem{Roh3} H. Roh, ``QCD Confinement Mechanism and $\Theta$ Vacuum:
Dynamical Spontaneous Symmetry Breaking,'' hep-th/0012261.
\bibitem{Namb} Y. Nambu, Phys. Rev. Lett. {\bf 4}, 380 (1960);
J. Goldstone, Nuovo Cimento {\bf 19}, 154 (1961).
\bibitem{Roh2} H. Roh, ``Quantum Weakdynamics as an $SU(3)_I$ Gauge Theory:
Grand Unification of Strong and Electroweak Interactions.'' hep-th/0101001.
\bibitem{Meis} W. Meissner and R. Ochsenfeld, Naturwiss. {\bf 21}, 787 (1933).
\bibitem{Aitc} See, e.g., I. J. R. Aitchison and A. J. G. Hey, {\em Gauge Theories in Particle Physics},
2nd Ed., pp 393 - 421, IOP Publishing Ltd. (1989).
\bibitem{Hubb} E. Hubble, Proc. Natl. Acad. Sci. (USA) {\bf 15}, 168 (1929).
\bibitem{Lind} See, e.g., A. Linde, {\em Particle Physics and Inflationary Cosmology}, pp 31 - 32, Harwood and Academic Pub. (1990).
\bibitem{Guth} A. H. Guth, Phys. Rev. {\bf D 23}, 347 (1981);
A. D. Linde, Phys. Lett. {\bf B 108}, 389 (1982);
A. Albrecht and P. J. Steinhardt, Phys. Rev. Lett. {\bf 48}, 1220 (1982).
\bibitem{Alta} I. S. Altarev et al., JETP Lett. {\bf 44}, 460 (1986);
K. F. Smith et al., Phys. Lett. {\bf B 234}, 191 (1990).
\bibitem{Chri} J. H. Christensen, J. W. Cronin, V. L. Fitch, and R. Turlay, Phys. Rev. Lett. {\bf 13}, 138 (1964).
\bibitem{Stei0} R. I. Steinberg, Ann. Rev. Astron. Astrophys. {\bf 14}, 339 (1976).
\bibitem{Berr} See, e.g., M. V. Berry, {\em Principles of Cosmology and Gravitation}, pp 74 - 88, Adam Hilger (1989).
\bibitem{Penz} A. A. Penzias and R. W. Wilson, Astrophys. J. {\bf 142}, 419 (1965).
\bibitem{Peeb} P. J. E. Peebles, D. N. Schramm, E. L. Turner, and R. G. Kron, Nature {\bf 352}, 769 (1991).
\bibitem{Arp} H. C. Arp, G. Burbidge, F. Hoyle, J. V. Narlikar, and N. C. Wickramansinghe, Nature {\bf 346}, 807 (1990).
\end{references}
\end{document}